\documentclass[aps, prl, reprint, floatfix, superscriptaddress, twocolumn, showpacs]{revtex4-1}
\usepackage[usenames,dvipsnames,svgnames,table]{xcolor}
\definecolor{HanRed}{RGB}{208, 16, 76}
\usepackage{hyperref}
\hypersetup
{
    colorlinks=true,
    citecolor=blue,
    linkcolor=HanRed,
    filecolor=magenta,
    urlcolor=cyan
}
\usepackage{changes}
\usepackage{silence}
\usepackage{natbib}
\usepackage{tikz}
\usepackage{textcomp}
\usepackage{amsmath}
\usepackage{amssymb}
\usepackage{bm}
\usepackage[normalem]{ulem}
\usepackage{soul}
\usepackage{siunitx}
\usepackage[T1]{fontenc}
\usepackage{times}
\usepackage{mathptmx}

\newcommand*{\Scale}[2][4]{\scalebox{#1}{$#2$}}%
\newcommand{\NFO}{NiFe$_\text{2}$O$_\text{4}$}

\newcommand{\REFF}[2]{~\ref{#1}\hyperref[#1]{(#2)}}
\PassOptionsToPackage{unicode}{hyperref}
\PassOptionsToPackage{naturalnames}{hyperref}

\begin{document}
\title{\textcolor{blue}{{Vectorial observation of the spin Seebeck effect in epitaxial NiFe$_\text{2}$O$_\text{4}$ thin films \mbox{with various magnetic anisotropy contributions}}}}
\author{Zhong Li}
\thanks{These authors contributed equally to this work.}
\affiliation{Center for Materials for Information Technology, The University of Alabama, Tuscaloosa, Alabama 35487, USA}
\affiliation{Department of Physics \& Astronomy, The University of Alabama, Tuscaloosa, Alabama 35487, USA}
\author{Jan Krieft}
\thanks{These authors contributed equally to this work.}
\affiliation{Center for Spinelectronic Materials and Devices, Department of Physics, Bielefeld University, Universit\"atsstra\ss e 25, 33615 Bielefeld, Germany}
\author{Amit Vikram Singh}
\affiliation{Center for Materials for Information Technology, The University of Alabama, Tuscaloosa, Alabama 35487, USA}
\author{Sudhir Regmi}
\affiliation{Center for Materials for Information Technology, The University of Alabama, Tuscaloosa, Alabama 35487, USA}
\affiliation{Department of Physics \& Astronomy, The University of Alabama, Tuscaloosa, Alabama 35487, USA}
\author{\mbox{Ankur Rastogi} }
\affiliation{Center for Materials for Information Technology, The University of Alabama, Tuscaloosa, Alabama 35487, USA}
\author{\mbox{Abhishek Srivastava}}
\affiliation{Center for Materials for Information Technology, The University of Alabama, Tuscaloosa, Alabama 35487, USA}
\affiliation{Department of Physics \& Astronomy, The University of Alabama, Tuscaloosa, Alabama 35487, USA}
\author{Zbigniew Galazka}
\affiliation{Leibniz-Institut f\"ur Kristallz\"uchtung, Max-Born-Str.\,2, 12489 Berlin, Germany}
\author{Tim Mewes}
\affiliation{Center for Materials for Information Technology, The University of Alabama, Tuscaloosa, Alabama 35487, USA}
\affiliation{Department of Physics \& Astronomy, The University of Alabama, Tuscaloosa, Alabama 35487, USA}
\author{Arunava Gupta}
\email[E-mail: ]{agupta@mint.ua.edu}
\affiliation{Center for Materials for Information Technology, The University of Alabama, Tuscaloosa, Alabama 35487, USA}
\author{Timo Kuschel}
\email[E-mail: ]{tkuschel@physik.uni-bielefeld.de}
\affiliation{Center for Spinelectronic Materials and Devices, Department of Physics, Bielefeld University, Universit\"atsstra\ss e 25, 33615 Bielefeld, Germany}
\date{\today}
\begin{abstract}
	We have developed a vectorial type of measurement for the spin Seebeck effect (SSE) in epitaxial NiFe$_\text{2}$O$_\text{4}$ thin films which have been grown by pulsed laser deposition on MgGa$_\text{2}$O$_\text{4}$ (MGO) with (001) and (011) orientation as well as CoGa$_\text{2}$O$_\text{4}$ (011) (CGO), thus varying the lattice mismatch and crystal orientation. We confirm that a large lattice mismatch leads to strain anisotropy in addition to the magnetocrystalline anisotropy in the thin films using vibrating sample magnetometry and ferromagnetic resonance measurements. Moreover, we show that the existence of a magnetic strain anisotropy in NiFe$_\text{2}$O$_\text{4}$ thin films significantly impacts the shape and magnitude of the magnetic-field-dependent SSE voltage loops. We further demonstrate that bidirectional field-dependent SSE voltage curves can be utilized to reveal the complete magnetization reversal process, which establishes a vectorial magnetometry technique based on a spin caloric effect.
\end{abstract}
\maketitle
The discovery of the spin Seebeck effect (SSE), the generation of a spin current from a thermal gradient, has attracted a lot of interest in the field of spin caloritronics in recent years~\cite{Uchida2008,Uchida2010b,Weiler2012,Meier2013a,Meier2015,uchida2016}. Pt/YIG (yttrium iron garnet, Y$_\text{3}$Fe$_\text{5}$O$_\text{12}$) is one of the most widely studied material systems in spin caloritronics \cite{Bauer2012} and magnon spintronics \cite{chumak2015} regarding the SSE, coherent spin pumping \cite{tserkovnyak2002}, spin Hall magnetoresistance (SMR) \cite{nakayama2013} and nonlocal magnon spin transport \cite{cornelissen2015long}. This material system has also been used to introduce the longitudinal spin Seebeck effect (LSSE), the generation of a spin current parallel to a temperature gradient that is usually aligned out-of-plane~\cite{Uchida2010b}. The spin current is converted to charge current in the Pt layer via the inverse spin Hall effect (ISHE)~\cite{saitoh2006}. In order to optimize the performance in LSSE experiments, a number of improvements have been investigated, including studies on the influence of YIG film thickness, temperature, applied magnetic field and interface between the spin-current-detecting material and YIG layers~\cite{Kehlberger2015,Uchida2015,Kikkawa2015,aqeel2014,qiu2015influence}. The spin Hall conductivity of the spin-current detector changes systematically in response to the number of $d$ electrons in the $4d$ and $5d$ transition metals \cite{morota2011indication}. Besides Pt, another material of this group with an effective spin-charge conversion and thus a sufficiently high spin-Hall angle is Pd \cite{ando2010inverse}, exhibiting highly transparent nonmagnetic/ferromagnetic material e.g. Pd/YIG interfaces \cite{ma2018spin} with strong
potential for spintronic applications \cite{tao2018self}.

NFO (nickel ferrite, \NFO) thin films, which have potential applications in high-frequency microwave and spintronics devices \cite{luders2006,klewe2014physical,hoppe2015,bougiatioti2017electrical}, have also been used in LSSE studies~\cite{Meier2013a,Meier2015,Shan2017a,Bougiatioti2017}.
Recent studies for NFO on lattice-matched substrates, such as MgGa$_\text{2}$O$_\text{4}$, report strongly improved magnetic properties which are comparable to YIG \cite{Singh2017}, as well as enhanced SSE and improved spin transport characteristics~\cite{shan2018}. The choice of substrate material and crystal cut can be used to taylor the magnetic anisotropy in the thin NFO film in order to study the impact of the anisotropy type on the SSE. Using a four-contact device, we present in this letter a vectorial magnetometry technique based on LSSE that is suitable to study magnetic anisotropies and magnetization reversal processes. We demonstrate the technique on NFO thin films grown on lattice-matched substrates with different crystal cuts and, thus, study different magnetic anisotropy combinations.

Within this Letter, the thermal generation of spin currents has been realized in NFO thin films deposited on MgGa$_\text{2}$O$_\text{4}$ (MGO) with (001) and (011) orientation as well as CoGa$_\text{2}$O$_\text{4}$ (011) (CGO), thus varying the lattice mismatch. Pd has been used as spin-current detector material. Using x-ray diffraction (XRD), we show that the lattice mismatches of NFO films on MGO substrates are larger than those on CGO substrates. A larger lattice mismatch leads to a higher magnetic strain anisotropy in NFO//MGO thin films, as confirmed via vibrating sample magnetometry (VSM) and ferromagnetic resonance (FMR) measurements. The vector detection technique has been used to study the influence of magnetic strain anisotropy in thin films on the SSE response. We find that the shapes of SSE voltage curves detected by two orthogonally aligned voltage probes measured as a function of the magnetic field strength and its orientation can vary significantly from each other due to the effect of magnetic strain anisotropy. While first attempts in this direction focused on detecting individual magnetic components separately~\cite{Kehlberger2014}, we demonstrate our simultaneous measurement method as a tool to study the magnetization reversal process using the SSE signals detected by the two perpendicularly aligned voltage probes. 
\begin{figure}[!htb]%
	\centering
	\includegraphics[width=1\linewidth]{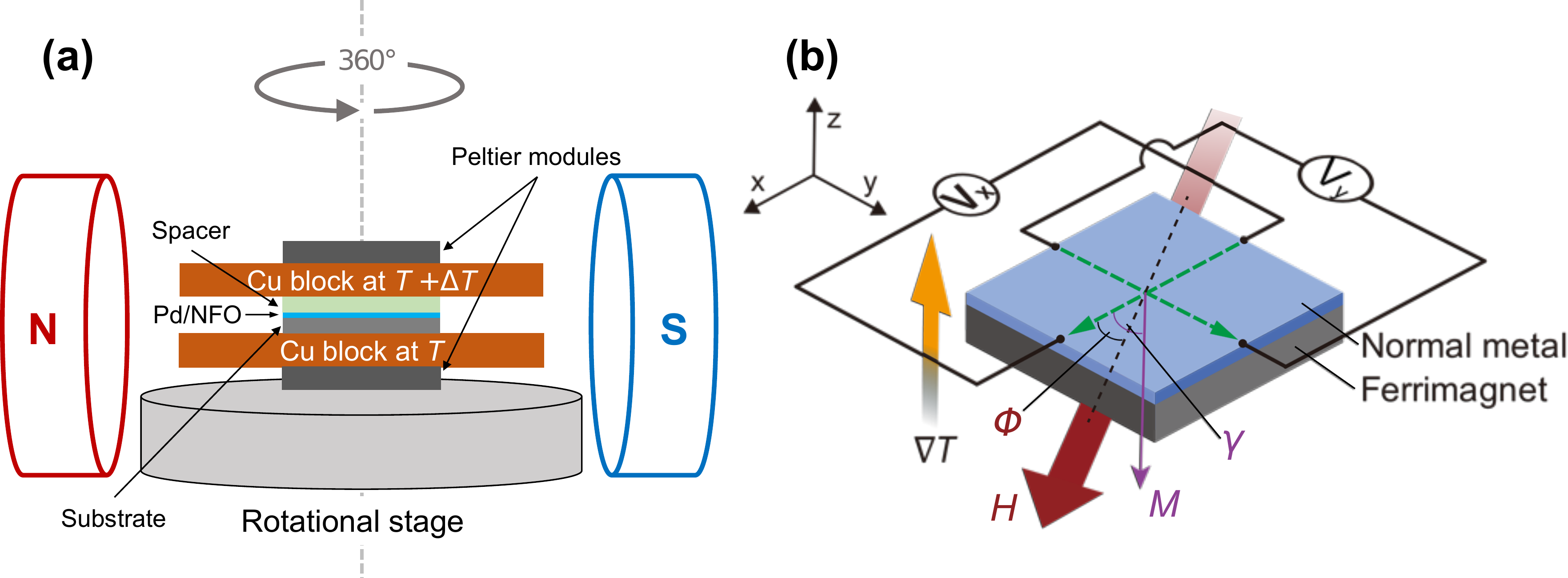}%
	\caption{(Color online) (a) LSSE setup. The sample is placed between two Cu blocks and a temperature gradient $\nabla$\textit{T} is applied. (b) The geometry for four-point vectorial LSSE measurements and definition of the external magnetic field and magnetization angles $\bm\phi$ and $\bm\gamma$.} %
	\label{Schematic}%
\end{figure}%

All films were grown using pulsed laser deposition (PLD) and structurally characterized using a Philips X$^\prime$Pert diffractometer with a Cu-K$_\alpha$ source, which helped to quantify the lattice mismatches for the different substrates (Supplementary Material (SM) II).
Magnetization hysteresis loops of the samples were measured by VSM in a PPMS$^{\text{\circledR}}$ DynaCool$^\text{{TM}}$ system. In the setup for LSSE measurements (Fig.\REFF{Schematic}{a}), the sample was placed between two Cu blocks. To carry out vectorial LSSE measurements (Fig.\REFF{Schematic}{b}), four Al wires were bonded on four points of the Pd layer so that they are located on two orthogonally aligned axes. The two contacts of each axis were connected to a separate nanovoltmeter (more details in SM I).

\begin{figure}[!htb]%
	\centering
	\includegraphics[width=1\linewidth]{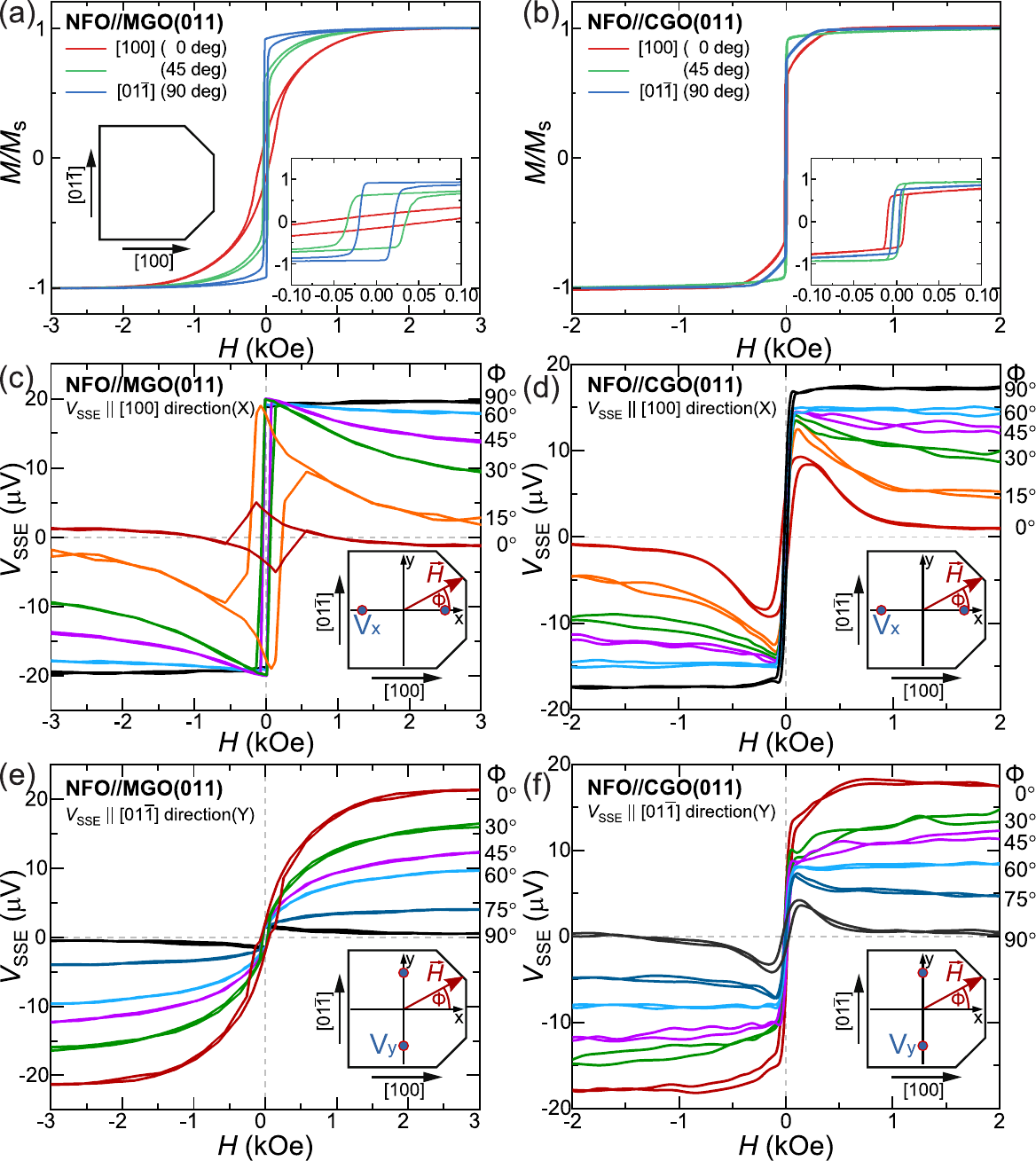}%
	\caption{(Color online) (a),(b) Normalized in-plane magnetization versus magnetic field for NFO//MGO (011) and NFO//CGO (011) (The insets are substrate orientation schematic and close-up figures). The magnetization is measured with an external in-plane magnetic field applied in three different magnetic field direction ($0^\circ$ [100], $45^\circ$ and $90^\circ$ [01$\bar{\text{1}}$]). (c)-(f) LSSE measurements at various angles $\bm\phi$ for (c),(e) NFO//MGO (011) and (d),(f) NFO//CGO (011) with voltage measured (c),(d) along the [100] (X) direction and (e),(f) along the [01$\bar{\text{1}}$] (Y) direction ($\Delta \textit{T} = 20$\,K).}%
	\label{FIG_VSM_SSE}%
\end{figure}

Magnetization measurements were performed by VSM on $\sim$450 nm thick NFO films grown on (011)- and (001)-oriented MGO and (011)-oriented CGO substrates to examine their magnetic in-plane anisotropy characteristics. While the results for NFO//MGO (011) and NFO//CGO (011) samples are presented in Figs.~2(a) and (b), we refer to the SM III for NFO//MGO (001). For the NFO//MGO (011) thin film, we observe a sharp switching of the magnetization when the external magnetic field is applied along the [01$\bar{\text{1}}$] direction (Fig.~\REFF{FIG_VSM_SSE}{a}). With the external magnetic field applied along the [100] direction, we observe magnetic hard axis type switching behavior with an anisotropy field of $\sim$1500 Oe. In addition, we measured the SSE voltage signal $V_{\text{LSSE}}$ of the sample along the two perpendicular directions (Fig.\REFF{Schematic}{b}). In the first configuration, the $V_{\text{LSSE}}$ signal is measured along the [100] direction (Fig.~\REFF{FIG_VSM_SSE}{c}), while the $V_{\text{LSSE}}$ signal is measured along the [01$\bar{\text{1}}$] direction in the second configuration (Fig.~\REFF{FIG_VSM_SSE}{e}).

In the first configuration (Fig.\REFF{FIG_VSM_SSE}{c}), when the magnetic field is applied along the [01$\bar{\text{1}}$] direction, the magnetization $\bm{M}$ of the NFO film is also aligned in the same direction. When the magnetic field direction changes polarity, the magnetization of the film $\bm{M}$ also switches into the opposite direction. This results in a sharp switching in the $V_{\text{LSSE}}$ signal ($\bm\phi=90^\circ$ in Fig.\REFF{FIG_VSM_SSE}{c}) and it is comparable to the corresponding magnetization measurement when the magnetic field is in the [01$\bar{\text{1}}$] direction (Fig.\REFF{FIG_VSM_SSE}{a}). In the next step, we changed the angle $\bm\phi$ of the external magnetic field with respect to the $x$ axis (insert in Fig.\REFF{FIG_VSM_SSE}{c}) in the range from 0$^\circ$ to 90$^\circ$. In saturation, the magnetization of the NFO film $\bm{M}$ is almost aligned along the direction of the external magnetic field for all $\bm\phi$ angles. The voltage generation (${V}_{\text{LSSE}} = {E}_{\text{LSSE}} \cdot d$) due to the ISHE is characterized by the projection of magnetization onto the [01$\bar{\text{1}}$] direction. The electric field ${E}_{\text{LSSE}}$ generated by the SSE is given by
\begin{equation}\label{eq: SSE}
	\bm{E}_{\text{LSSE}} \propto \bm{J}_\text{s} \times \bm{\sigma},
\end{equation}
where $\bm{J}_\text{s}$ is the thermally induced spin current which is parallel to the $\nabla$\textit{T} in $z$ direction, and $\bm{\sigma}$ is the spin polarization vector which is aligned along $\bm{M}$. Thus, from Eq.~\eqref{eq: SSE} we can conclude that
\begin{equation}\label{eq: SSEcomponent}
	{V}_{\text{x}} \propto \bm{\sigma}_\text{y} \propto M_\text{y} \; , \; {V}_{\text{y}} \propto \bm{\sigma}_\text{x} \propto M_\text{x}.
\end{equation}
With increasing the angle $\bm\phi$ between the external magnetic field and the [100] direction along the $x$ axis, the saturation voltage ${V}_{\text{x}}$ increases in correspondence with a factor of $\sin{\bm\phi}$ due to the cross product of the ISHE~\cite{saitoh2006}.
When the projection of the magnetization in the [01$\bar{\text{1}}$] direction $M_\text{y}$ increases (decreases) due to the increase (decrease) of the external field, the measured voltage signal of the ISHE also increases (decreases).
At zero magnetic field the magnetization can rotate completely into the magnetic easy axis or partially (or fully) switch into another magnetic easy axis reducing the projection $M_\text{y}$ and the corresponding voltage ${V}_{\text{x}}$.
When increasing the external field into opposite direction the voltage usually changes sign and the coherent rotation of the  magnetization out of the magnetic easy axis along the [01$\bar{\text{1}}$] direction \mbox{($\bm\phi$ $<$ 90$^{\circ}$)} is accompanied by a decrease of the absolute voltage value (Fig.\REFF{FIG_VSM_SSE}{c}).
For $\bm\phi$ = 0$^{\circ}$, the saturation voltage is nearly zero as we expect from the ISHE. The small residual voltage signal in saturation could be explained by a slight misalignment of voltage contacts along the [100] direction or with an alignment of the magnetization that is not fully saturated.

In the second configuration (Fig.\REFF{FIG_VSM_SSE}{e}), the voltage contacts along the [01$\bar{\text{1}}$] (Y) direction are used. When the external magnetic field is applied along the [100] direction to complete the typical LSSE configuration (i.e.~$\bm\phi$ = 0$^{\circ}$), we observe a similar voltage value in magnetic saturation as compared to the previous configuration in Fig.\REFF{FIG_VSM_SSE}{c}). When the external magnetic field decreases, the LSSE voltage signal does not show a sharp switching, but follows the magnetization measurement in the [100] direction with low remanence (in Fig.~\REFF{FIG_VSM_SSE}{a}), which significantly differs from that in the first configuration. While the projection of the magnetization onto the [100] direction $M_\text{x}$ increases (decreases) monotonically until the magnetization switches, the LSSE voltage ${V}_{\text{y}}$ also increases (decreases). This is comparable to the first configuration in Fig.\REFF{FIG_VSM_SSE}{c}, detecting the same switching events while being sensitive to orthogonal projections of the magnetization vector.
Therefore, the LSSE measurements provide a promising alternative compared to established optical measurement methods to determine both in-plane components of the magnetization during field reversal.

Figure\REFF{FIG_VSM_SSE}{b} shows the VSM results for the NFO//CGO (011) thin film as a comparison with the NFO//MGO (011) sample.
The magnetization curves for different external field directions still indicate magnetic easy axis behavior in [01$\bar{\text{1}}$] direction as well as magnetic hard axis characteristics in [100] direction. However, the differences between easy-axis and hard-axis loops are not that pronounced as seen for the NFO//MGO (011) sample. 
We also performed SSE measurements on this sample. As shown in Figs.\REFF{FIG_VSM_SSE}{d} and\REFF{FIG_VSM_SSE}{f}, we find that the SSE measurements along these two perpendicular directions are very similar, which is consistent with the previous magnetization results in Fig.\REFF{FIG_VSM_SSE}{b}.

\begin{figure}[!b]%
	\centering
	\includegraphics[width=0.5\linewidth]{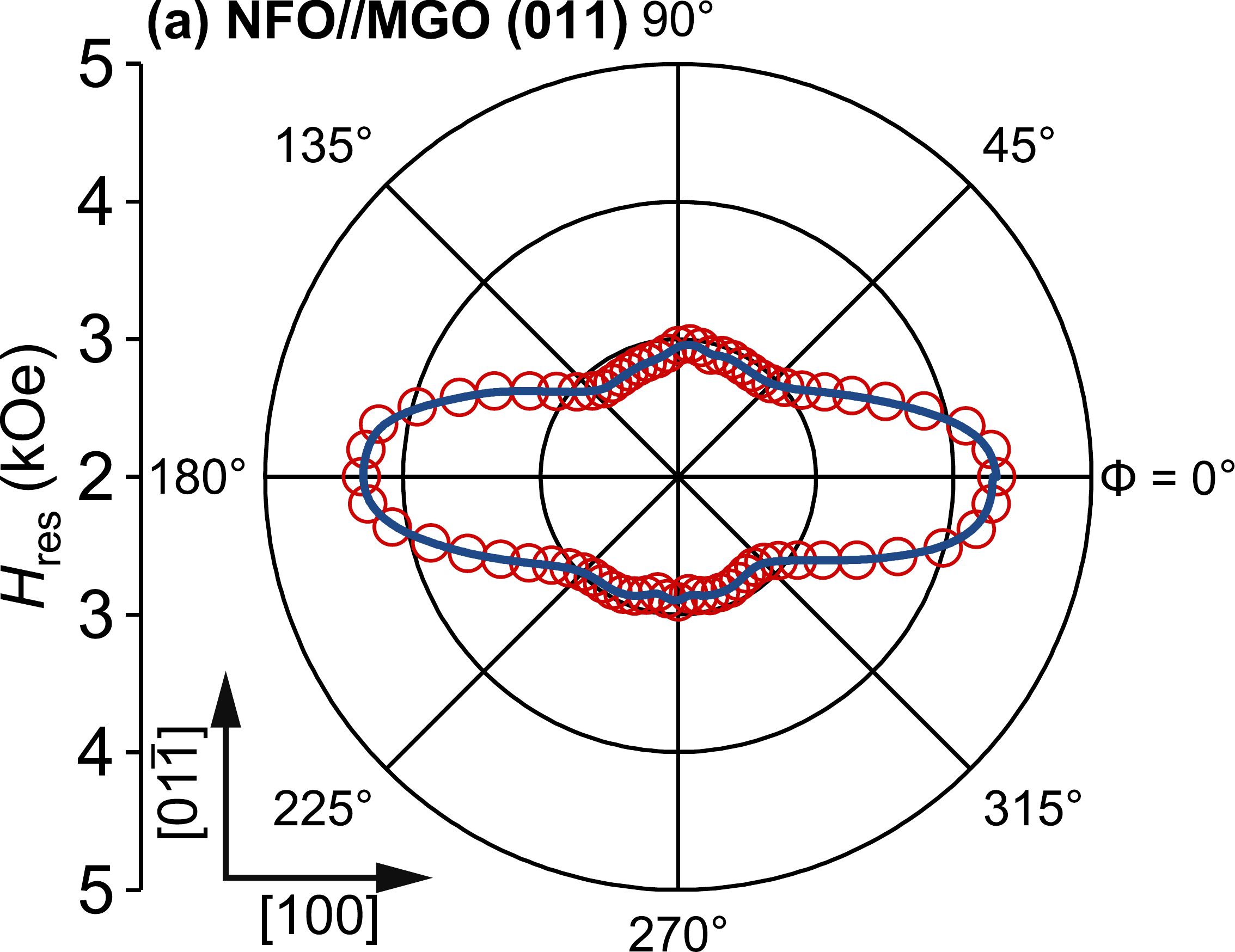}%
	\includegraphics[width=0.5\linewidth]{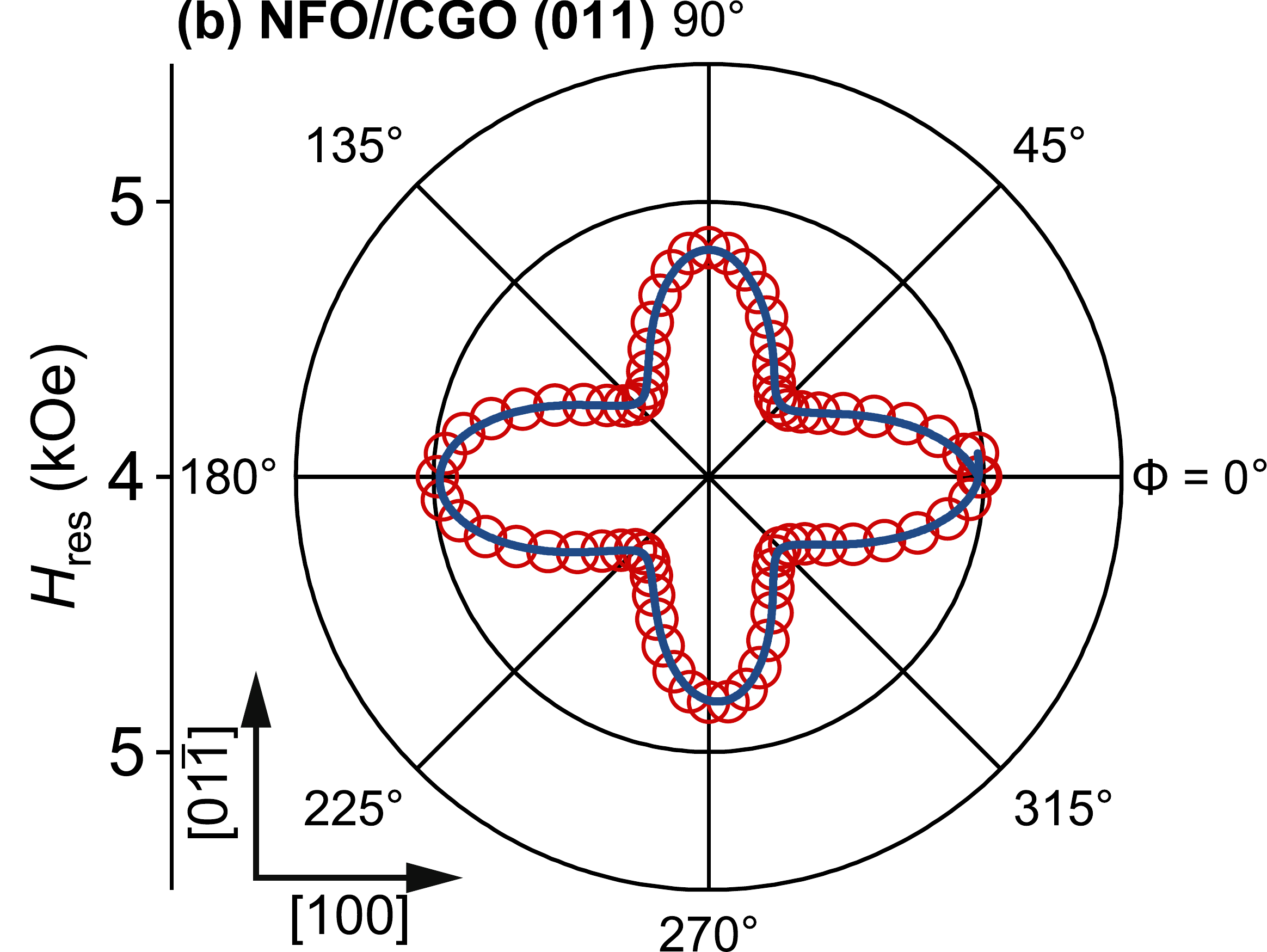}%
	\caption{(Color online) In-plane angular dependence of resonance field, $H_{\text{res}}$, at 20 GHz for (a) NFO//MGO (011) and (b) NFO//CGO~(011).} %
	\label{FIG_FMR}%
\end{figure}%

We have further carried out in-plane angular-dependent FMR measurements that identify the presence of a fourfold combined with a strong uniaxial anisotropy for NFO//MGO (011) thin films. The results are shown in Fig.\REFF{FIG_FMR}{a}, with overall in-plane magnetic easy axis along the [01$\bar{\text{1}}$] direction and the magnetic hard axis along the [100] direction.
On the other hand, for NFO//CGO (011) thin films, the fourfold anisotropy is dominant and only a weak uniaxial anisotropy is present, resulting in a biaxial anisotropy with a more and less hard axis in 0$^\circ$ and 90$^\circ$ direction, as shown in Fig.\REFF{FIG_FMR}{b}. Thus, we conclude that in addition to the anisotropy landscape modified by the crystal cut a strain anisotropy affects the overall strength and direction of magnetic easy and hard axes. While we observe this anisotropy modification with various extent for NFO//MGO (011) and NFO//CGO (011) samples, the NFO//MGO (001) control sample exhibits a pure four-fold magnetocrystalline anisotropy. For the FMR results of this sample and a deeper fit analysis of the FMR measurements presented here see SM VI.

The reversal process of the magnetization vector depends on the magnetic anisotropy and the direction of the external magnetic field. The data used to display the LSSE hysteresis loops (in Figs.\REFF{FIG_VSM_SSE}{c} -\REFF{FIG_VSM_SSE}{f}) can now be used in order to reconstruct the magnetization reversal process, utilizing both projections of the in-plane magnetization vector on the [100]  and [01$\bar{\text{1}}$] directions from the vector LSSE measurement ($M_\text{x}$ and $M_\text{y}$).  From Eq.~\eqref{eq: SSEcomponent}, we can assume that
\begin{equation}\label{eq: Mcomponent}
	{V}_\text{x} = A_\text{x}\cdot M_\text{y} \; ,\; {V}_\text{y} = A_\text{y}\cdot M_\text{x} ,
\end{equation}
where $A_\text{x}$ and $A_\text{y}$ are material and setup parameters. A detailed analysis of the voltage $V_\text{SSE}$ anisotropy is presented in SM IV.

The LSSE saturation voltages $V_\text{x}$ and $V_\text{y}$ in relation to the azimuthal angle of the magnetization show no significant anisotropy yet in principle the magnetic transport and the magnitude of the generated SSE voltages can depend on the relative orientation to the NFO crystal axes. Notably in SMR measurements a current direction anisotropy can be observed in NFO thin films where the precise origin of the current anisotropy is still up for debate \cite{althammer2018current}. In this SMR and in our SSE experiment the spin current is always generated in out-of-plane direction propagating along the same crystallographic axis and is therefore independent from the voltage detection direction. Still, an anisotropy may originate from anisotropic SHE or spin mixing conductance as discussed by Althammer \textit{et al.} \cite{althammer2018current}. However, as presented in Fig.\REFF{FIG_VSM_SSE}{c-f} the amplitude of $V_\text{SSE}$ at maximum field measured parallel to [100] direction is comparable to the voltage measured along the [01$\bar{\text{1}}$] direction for both NFO//MGO (011) as well as NFO//CGO (011). This is a strong indication that there is no significant anisotropy of the ISHE e.g. by strain similar to the reported immunity to electrostrain of the spin-current transport and spin-charge conversion in Pt/YIG by Wang \textit{et al.} due to robust magnon diffusion length and spin-mixing conductance at the interface \cite{wang2018voltage}.

\begin{figure}[!tb]%
	\centering
	\includegraphics[width=1\linewidth]{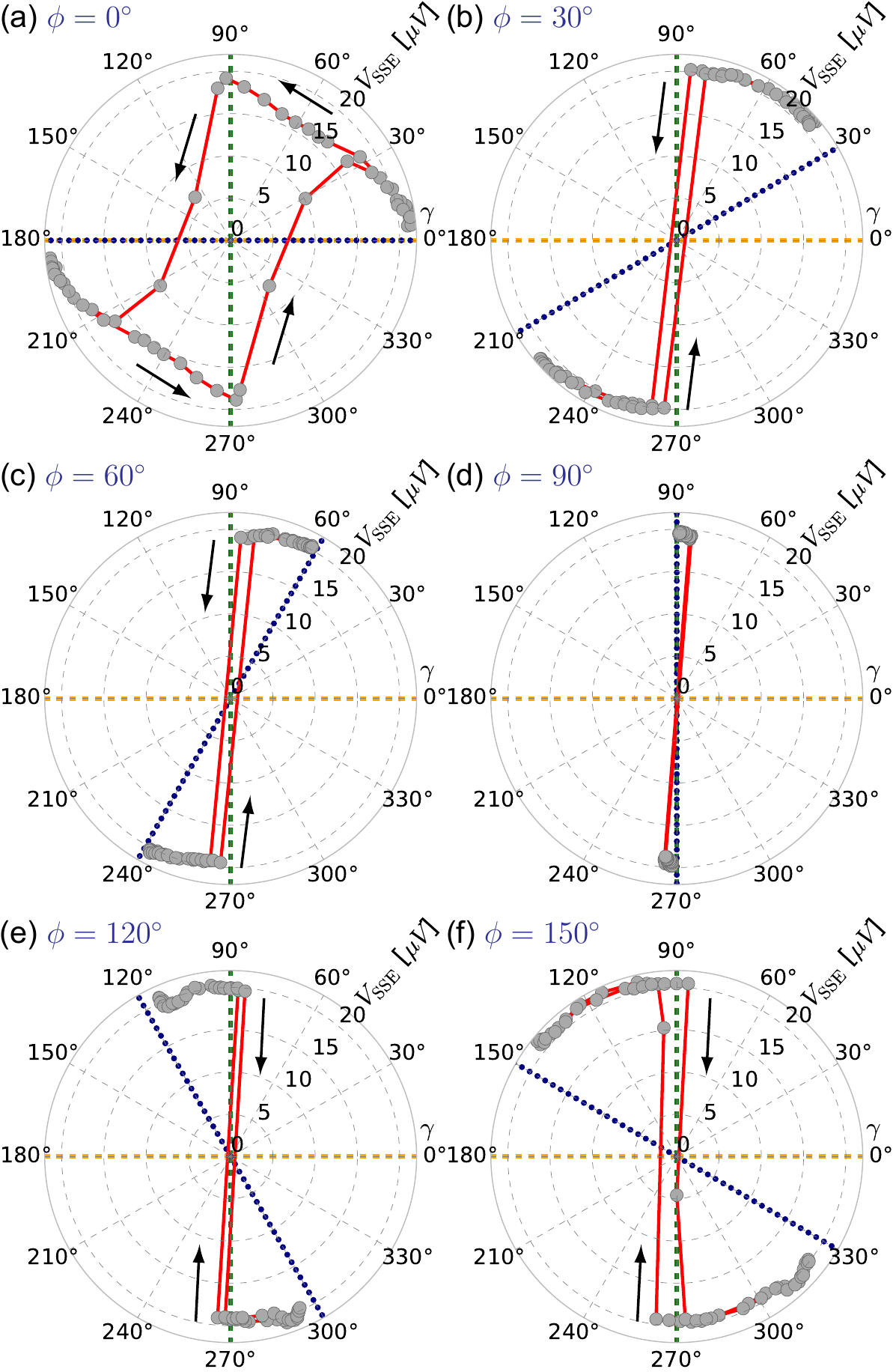}
	\caption{(Color online) The NFO//MGO (011) reversal processes of the magnetization vector for angles of the external magnetic field (blue dotted line) $\bm\phi$ from 0$^\circ$ to 150$^\circ$ ((a) - (f)) relative to the [100] in-plane direction (inset Fig.\REFF{FIG_VSM_SSE}{a}). The magnitude of the magnetization vector given by $|V|$ of the combined LSSE measurements in $\SI{}{\micro\volt}$  is plotted against the rotation angle $\bm{\gamma}$ of the magnetization vector. The magnetic easy (green) and magnetic hard axes (orange) are marked by dashed lines.
	} %
	\label{FIG_185_reversal}%
\end{figure}%

Subsequently, we can use the assumption $A_\text{x} \approx A_\text{y}$  to calculate other quantities like the magnetization $\bm{M}$,
\begin{equation}\label{eq: Mtotal}
	\Scale[0.93]{{M} = \sqrt{(M_\text{x})^2 + (M_\text{y})^2}=\sqrt{(\frac{V_\text{y}}{A_\text{y}})^2 + (\frac{V_\text{x}}{A_\text{x}})^2}\approx \frac{1}{A_\text{x}}\cdot\sqrt{V_\text{y}^2 + V_\text{x}^2}}.
\end{equation}
We calculate the in-plane orientation of the magnetization vector based on the magnitude $|V|=\sqrt{{V_\text{x}}^2 + {V_\text{y}}^2}$, and the azimuthal magnetization angle $\bm{\gamma}$ can be expressed as
\begin{equation}\label{eq: gamma}
	\bm{\gamma} =\arctan(\frac{M_\text{y}}{M_\text{x}})=\arctan(\frac{V_\text{x}}{V_\text{y}}\cdot \frac{A_\text{y}}{A_\text{x}})\approx\arctan(\frac{V_\text{x}}{V_\text{y}}).
\end{equation}
This can now be applied to create polar plots of the magnetization vector length versus $\bm{\gamma}$. The plot compilation shown in Fig.~\ref{FIG_185_reversal} visualizes the NFO//MGO (011) reversal processes of the magnetization vector for the selected external magnetic field directions $\bm\phi$, defined as shown in Fig.~\ref{Schematic}. These polar plots represent the progress of the resulting magnetization vector state for each magnetic field strength during the stepwise change of the external magnetic field direction with respect to the sample orientation.
This analysis is based on the magnetization measurements shown in Fig.~\ref{FIG_VSM_SSE}. A defined magnetic domain state exists at the beginning of each reversal process, since the external magnetic field strength is sufficient to mainly saturate the magnetization orientation along the external magnetic field direction. The subplot Fig.\REFF{FIG_185_reversal}{a} exemplifies a reversal process when the external magnetic field is applied along the magnetic hard axis of the NFO//MGO (011) sample ($\bm\phi = 0^\circ$) according to the geometry defined in Fig.\REFF{Schematic}{b}.
A saturating external magnetic field yields a maximum in ${V}_{\text{y}}$ (from Eq.~\eqref{eq: SSEcomponent}) and represents the starting point of this reversal process as seen on the right side of Fig.\REFF{FIG_185_reversal}{a}. The application of an external field along or close to a magnetic hard axis induces domain splitting if the external magnetic field is decreased as indicated by the reducing magnitude of the magnetization vector. When the external magnetic field is further reduced (still applied along $\bm\phi = 0^\circ$), the magnetic moments of those domains switch or rotate towards the magnetic easy axis along [01$\bar{\text{1}}$] direction, indicated by the black arrow at the top of Fig.\REFF{FIG_185_reversal}{a}. The preferred direction is defined by the slight misalignment of the external magnetic field relative to the magnetic hard axis [100] direction of the thin film. With an increasing opposite external field completing the first branch of the LSSE hysteresis loop, the magnetization rotates back to the external field direction.

However, if the external magnetic field is not applied close to a magnetic hard axis the system essentially follows the coherent rotation model without transforming into multi-domains. Here, the length of the magnetization vector is constant from saturation to remanence. For an external magnetic field along the strong magnetic easy axis direction ($\bm\phi = 90^\circ$, Fig.\REFF{FIG_185_reversal}{d}), we can observe a simple switching of the magnetization direction by 180$^\circ$, as expected along a magnetic easy axis direction where the LSSE hysteresis in the standard measurement geometry shows a very small coercive field and high remanence. We furthermore used this vector LSSE technique to visualize the magnetization reversal process of corresponding samples showing a pure fourfold and a fourfold plus weak twofold anisotropy. The vector LSSE measurement results for those NFO//MGO~(001) and NFO//CGO~(011) samples can be found in the SM V.

The vectorial magnetization technique based on SSE is using electrical detection compared to magnetooptic instruments extracting the in-plane magnetization components by different combinations of magnetooptic Kerr effects (MOKE)~\cite{Daboo1993,mewes2004separation,
kuschel2011vectorial}.
In contrast to the magnetooptic techniques, our vectorial SSE approach is also applicable for magnetic materials that are not amenable to magnetooptic detection due to a vanishingly small Kerr rotation especially with standard vector MOKE techniques in the visible range of light such as ferrites, YIG~\cite{kehlberger2015enhanced} and in particular NFO. This illustrates the need for a wider range of applicable techniques for the probe of magnetization reversals in the expanding field of spin caloritronics, spin orbitronics, and beyond. 
Still, this technique offers room for improvement and simplification to allow for a broader applicability. For example, it is possible to implement lock-in techniques or replace the bonding process by designing an insulating spacer material pre-coated with defined voltage contacts to increase precision and practicability.

In conclusion, we have experimentally found that the lattice mismatch between NFO film deposited on isostructural spinel CGO (011) and MGO substrates of two different orientations ((001) and (011)) results in varying magnetic strain anisotropy as determined from VSM and FMR measurements. The strain anisotropy significantly influences the shape of the LSSE voltage hysteresis loop measured as a function of the external magnetic field.  Based on vector measurement of the LSSE, we show that the complete reversal process of the magnetization vector can be determined, demonstrating an alternative to study the magnetization reversal process of thin films based on SSE and ISHE voltage detection.

This work was supported by NSF ECCS Grant No.\,1509875 and the Deutsche Forschungsgemeinschaft (DFG) within the priority program Spin Caloric Transport (SPP 1538), Grant No.\,KU 3271/1-1. The MGO and CGO substrates were prepared from bulk crystals obtained by the Czochralski method at Leibniz-Institut f\"ur Kristallz\"uchtung, Berlin, Germany~\cite{Singh2017,galazka2015}. We finally thank Matthias Althammer and Weiwei Lin for additional input during the revision process of this letter.

\bibliographystyle{apsrev4-1}

\bibliography{bibliography}

\end{document}